\documentclass[twocolumn,pra,aps,superscriptaddress,floatfix]{revtex4}
\usepackage{amssymb}
\usepackage{amsfonts}
\usepackage{amsmath}
\usepackage{graphicx}

\begin{document}

\title{Quantum distinguishability measures: projectors vs. states
maximization }
\date{\today }
\author{Adri\'an A. Budini}
\affiliation{Consejo Nacional de Investigaciones Cient\'{\i}ficas y T\'{e}cnicas
(CONICET), Centro At\'{o}mico Bariloche, Avenida E. Bustillo Km 9.5, (8400)
Bariloche, Argentina, and Universidad Tecnol\'{o}gica Nacional (UTN-FRC),
Fanny Newbery 111, (8400) Bariloche, Argentina}
\author{Ruynet L. de Matos Filho}
\affiliation{Instituto de F\'{\i}sica, Universidade Federal do Rio de Janeiro, Caixa
Postal 68528, Rio de Janeiro, 21941-972, Brazil}
\author{Marcelo F. Santos}
\affiliation{Instituto de F\'{\i}sica, Universidade Federal do Rio de Janeiro, Caixa
Postal 68528, Rio de Janeiro, 21941-972, Brazil}

\begin{abstract}
The distinguishability between two quantum states can be defined in terms of
their trace distance. The operational meaning of this definition involves a
maximization over measurement projectors. Here we introduce an alternative
definition of distinguishability which, instead of projectors, is based on
maximization over normalized states (density matrices). It is shown that
this procedure leads to a distance (between two states) that, in contrast to
the usual approach based on a 1-norm, is based on an infinite-norm.
Properties such as convexity, monotonicity, and invariance under unitary
transformations are fulfilled. Equivalent operational implementations based
on maximization over classical probabilities and hypothesis testing
scenarios are also established. When considering the action of completely
positive transformations contractivity is only granted for unital maps. This
feature allows us to introduce a measure of the quantumness of non-unital
maps that can be written in terms of the proposed distinguishability measure
and corresponds to the maximal possible deviation from contractivity.
Particular examples sustain the main results and conclusions.
\end{abstract}

\maketitle

\section{Introduction}

Measuring the distinguishability between two quantum states is a central
ingredient when evaluating the performance of any quantum information
protocol. A solid basis of proposals and results have been developed in the
last years~\cite{nielsen,petz,fucks,chefles,wildeRep}. Nevertheless, due to
its relevance, this issue has been periodically reviewed and still remains
as an active area of research \cite%
{bruckner,bagan,Zyco,lamberti,audenaert,puchal,wilde,wang,agarwal}.

An usual and standard definition of distinguishability\ relies on the
following expressions~\cite{nielsen,petz,fucks,chefles,wildeRep}. Given two
quantum states $\rho _{A}$ and $\rho _{B}$ in an arbitrary Hilbert space,
their distinguishability is defined as%
\begin{equation}
D_{\Pi }(\rho _{A},\rho _{B})\equiv \max_{\{\Pi \}}\left\vert \mathrm{Tr}%
[\Pi (\rho _{A}-\rho _{B})]\right\vert .  \label{MaxProDef}
\end{equation}%
Here, $\mathrm{Tr}[\cdots ]$ is the trace operation. Maximization is
performed over arbitrary \textit{projectors}, $\Pi =\Pi ^{n}.$ In general,
these projectors may have an arbitrary rank (equal or greater than one). It
is well known that the operational definition of $D_{\Pi }(\rho _{A},\rho
_{B})$ is equivalent to the expression~\cite{nielsen}%
\begin{equation}
D_{\Pi }(\rho _{A},\rho _{B})=\frac{1}{2}\mathrm{Tr}|\rho _{A}-\rho _{B}|.
\label{DPi}
\end{equation}%
Hence, $D_{\Pi }(\rho _{A},\rho _{B})$ corresponds to the \textit{trace
distance} between the states $\rho _{A}$ and $\rho _{B}.$

Motivated by recent advances in the definition of environment quantumness in
open quantum systems~\cite%
{nori,franco,ChenChen,fata,lika,Szanko,sun,Liu,maps,Dq}, the main goal of
this paper is to introduce an alternative definition of quantum
distinguishability, providing in addition a full characterization of its
properties. While the standard approach~(\ref{MaxProDef}) relies on
maximization over projectors, here we propose to replace projectors with
normalized states, that is, density matrices.

We find that this alternative definition can be related to a distance based
on an infinite-norm. In contrast, the trace norm is related to a 1-norm~\cite%
{petz}. In addition we prove that some general properties hold in the
alternative approach. The proposed distinguishability measure is a metric on
the space of density matrices. Furthermore, it is convex on both entries.
Monotonicity and invariance under unitary transformations are also
fulfilled. Complementarily, we show that equivalent implementations can be
defined in terms of maximization over classical probabilities and hypothesis
testing scenarios~\cite{wildeRep,helstrom}. We also find the conditions
under which the state-based and projector-based distinguishability measures
are equal.

Added to the intrinsic theoretical and practical interest of the previous
results, we find that the alternative definition, in contrast to the
standard approach [Eqs.~(\ref{MaxProDef}) and~(\ref{DPi})], allows to
quantify departures from classicality of open quantum dynamics. This
quantum-classical border~\cite%
{nori,franco,ChenChen,fata,lika,Szanko,sun,Liu,maps,Dq} is studied by
considering the action of completely positive maps. Consistently with the
results of Ref.~\cite{ruskai}, we find that, in general, contractivity does
not hold here. Hence, the distance between the output states could increase
with respect to the distance between the input states. The specific states
and maps that lead to maximal violation of contractivity are explicitly
stated. These results provide the basis for defining a measure that
quantifies the quantumness of dissipative (non-unital) open system dynamics.
Furthermore, a close relationship with recently proposed measures of
environment quantumness~\cite{Dq} emerges from these analyses.

The manuscript is outlined as follows. In Sec. II we introduce the
distinguishability measure based on maximization over states. Its
relationship with an infinite norm is demonstrated. Equivalent operational
implementations such as maximization over classical probabilities and
hypothesis testing scenario are established. Furthermore, we compare the
projector and state-based measures establishing the conditions under which
they are equal. In Sec. III we study its properties when considering the
action of completely positive maps. In Sec. IV we study some specific
examples of distance between states and quantum maps. In Sec. V we provide
the Conclusions. Extra related results are provided in the Appendices.

\section{Distinguishability measure based on maximization over states}

Here, we introduce an alternative definition of distinguishability. In
contrast to a maximization over projectors [Eq.~(\ref{MaxProDef})] it
emerges from a maximization over states. Given two quantum states $\rho _{A}$
and $\rho _{B}$ $(\mathrm{Tr}[\rho _{A}]=\mathrm{Tr}[\rho _{B}]=1)$ it reads%
\begin{equation}
D_{\rho }(\rho _{A},\rho _{B})\equiv \max_{\{\rho \}}\left\vert \mathrm{Tr}%
[\rho (\rho _{A}-\rho _{B})]\right\vert ,  \label{DrhoDefinition}
\end{equation}%
where $\rho $ is an arbitrary density matrix, $\mathrm{Tr}[\rho ]=1.$ We
notice that, being a state, $\rho $ has positive eigenvalues in the interval 
$[0,1].$ Furthermore, similarly to $D_{\Pi },$ $D_{\rho }$ is also a
dimensionless quantity.

The definition~(\ref{DrhoDefinition}) can be read as a maximization over
states $\rho $ of the expectation value of the \textquotedblleft Hermitian
operator\textquotedblright\ $(\rho _{A}-\rho _{B}).$ In Appendix~\ref%
{MaxAverage} we provide a general solution to this problem (arbitrary
operator $A$). Introducing the eigenvalues and eigenvectors associated to $%
(\rho _{A}-\rho _{B}),$%
\begin{equation}
(\rho _{A}-\rho _{B})|i\rangle =\zeta _{i}|i\rangle ,  \label{Eigen}
\end{equation}%
the maximization in Eq.~(\ref{DrhoDefinition}) leads to [see Eq.~(\ref%
{LamMax})] 
\begin{equation}
D_{\rho }(\rho _{A},\rho _{B})=\max_{\{i\}}\{|\zeta _{i}|\}.
\label{DRhoEigenSol}
\end{equation}%
Hence, $D_{\rho }(\rho _{A},\rho _{B})$ corresponds to the eigenvalue of $%
(\rho _{A}-\rho _{B})$ with maximal absolute value. In contrast, notice that
the projector-based definition [Eq.~(\ref{DPi})] can be written as $D_{\Pi
}(\rho _{A},\rho _{B})=(1/2)\sum_{i}|\zeta _{i}|.$ On the other hand, the
state $\rho $ that solves the maximization in Eq.~(\ref{DrhoDefinition}),
while in general not unique (see Appendix~\ref{MaxAverage}), can always be
chosen as%
\begin{equation}
\rho =|i_{\max }\rangle \langle i_{\max }|,  \label{RhoMax}
\end{equation}%
where $|i_{\max }\rangle $ is the eigenstate of $(\rho _{A}-\rho _{B})$
associated to the eigenvalue with maximal absolute value, that is, $%
\max_{\{i\}}\{|\zeta _{i}|\}.$

In order to understand the difference between $D_{\rho }(\rho _{A},\rho
_{B}) $ and $D_{\Pi }(\rho _{A},\rho _{B})$ we notice that Eq.~(\ref%
{DRhoEigenSol}) can be written in the alternative way%
\begin{equation}
D_{\rho }(\rho _{A},\rho _{B})=\lim_{\alpha \rightarrow \infty }\sqrt[\alpha 
]{\mathrm{Tr}|\rho _{A}-\rho _{B}|^{\alpha }}.  \label{AlfaLimit}
\end{equation}%
This expression allows to read $D_{\rho }(\rho _{A},\rho _{B})$ as a
distance between states based on a infinite-norm while $D_{\Pi }(\rho
_{A},\rho _{B})$ [Eq.~(\ref{DPi})] is a distance based on a 1-norm [given an
operator $A,$ its $\alpha $-norm $(\alpha \geq 1)$ is given by $|A|_{\alpha
}=\sqrt[\alpha ]{\mathrm{Tr}|A|^{\alpha }}].$

The proposed distinguishability measure is defined by Eq.~(\ref%
{DrhoDefinition}), whose explicit calculation is solved by Eq.~(\ref%
{DRhoEigenSol}). In Appendix~\ref{PropertiesD} we demonstrate that $D_{\rho
}(\rho _{A},\rho _{B})$ fulfills some general properties. In particular, it
is shown that it defines a metric in the space of states, it is convex in
both entries and monotonicity for bipartite systems and invariance under
unitary transformations are also corroborated.

\subsection{Equivalent operational interpretations}

Below we study different equivalent operational interpretations of $D_{\rho
}.$

\subsubsection{Maximization in terms of probabilities}

Let $\{|k\rangle \}$ be the basis where an \textit{arbitrary} state $\rho $\
is diagonal, $\rho =\sum_{k}p_{k}|k\rangle \langle k|.$ Given two quantum
states $\rho _{A}$ and $\rho _{B}$ define $p_{A}^{(k)}\equiv \langle k|\rho
_{A}|k\rangle ,$ and $p_{B}^{(k)}\equiv \langle k|\rho _{B}|k\rangle .$
Then, the distinguishability measure [Eq.~(\ref{DrhoDefinition})] can
alternatively be written as%
\begin{equation}
D_{\rho }(\rho _{A},\rho _{B})=\max_{\{\rho \}}D_{c}(p_{A},p_{B}),
\label{MaxClassical}
\end{equation}%
where the maximization is over all possible states $\{\rho \}.$ With $%
p_{A}\equiv \{p_{A}^{(k)}\}$ and $p_{B}\equiv \{p_{B}^{(k)}\}$ we denote
both sets of probabilities. Their\ distinguishability is%
\begin{equation}
D_{c}(p_{A},p_{B})\equiv \max_{\{k\}}\{|p_{A}^{(k)}-p_{B}^{(k)}|\}.
\label{DC}
\end{equation}

We notice that Eq.~(\ref{MaxClassical}) implies that $D_{\rho }(\rho
_{A},\rho _{B})$ is the distinguishability $D_{c}(p_{A},p_{B})$\ between
probabilities maximized over all possible states $\rho .$ A similar result
is valid for $D_{\Pi }(\rho _{A},\rho _{B})$~\cite{nielsen}, but where the
probabilities are defined in terms of an arbitrary positive operator value
measure~\cite{POVM}.

\textit{Demonstration}: Below we demonstrate the validity of the operational
representation defined by Eqs.~(\ref{MaxClassical}) and~(\ref{DC}). By using
the explicit expressions of $p_{A}^{(k)}$ and $p_{B}^{(k)},$ it is possible
to rewrite $D_{c}(p_{A},p_{B})$ as%
\begin{equation}
D_{c}(p_{A},p_{B})=\max_{\{k\}}\{\left\vert \langle k|(\rho _{A}-\rho
_{B})|k\rangle \right\vert \}.
\end{equation}%
From Eq.~(\ref{Eigen}) we write%
\begin{equation}
(\rho _{A}-\rho _{B})=\sum_{i}\zeta _{i}\ |i\rangle \langle i|,
\label{DiferenciaDiagonal}
\end{equation}%
where $\{|i\rangle \}$ is the basis where $(\rho _{A}-\rho _{B})$\ is a
diagonal matrix. Hence, the previous expression becomes%
\begin{eqnarray*}
D_{c}(p_{A},p_{B}) &=&\max_{\{k\}}\Big{\{}\Big{|}\sum_{i}\zeta _{i}\
|\langle k|i\rangle |^{2}\Big{|}\Big{\}}, \\
&\leq &\max_{\{k\}}\Big{\{}\sum_{i}|\zeta _{i}|\ |\langle k|i\rangle |^{2}%
\Big{\}}, \\
&\leq &\Big{(}\max_{\{i\}}\{|\zeta _{i}|\}\Big{)}\
\max_{\{k\}}\sum_{i}|\langle i|k\rangle |^{2} \\
&=&\max_{\{i\}}\{|\zeta _{i}|\}=D_{\rho }(\rho _{A},\rho _{B}),
\end{eqnarray*}%
which demonstrates Eq.~(\ref{MaxClassical}). In fact, the equality is
achieved when the basis $\{|k\rangle \}$ where the state $\rho $ is diagonal
is the same basis $\{|i\rangle \}$ where $(\rho _{A}-\rho _{B})$ is a
diagonal operator.

\subsubsection{Hypothesis testing scenario}

Here we demonstrate that under an appropriate constraint $D_{\rho }$ plays
the same role that $D_{\Pi }$ in a \textquotedblleft hypothesis-testing
scenario\textquotedblright ~\cite{helstrom,wildeRep}.\ Let Alice prepare two
quantum states $\rho _{1}$ and $\rho _{0},$ each one with probability $1/2.$
Bob can perform a binary \textquotedblleft positive operator value
measure\textquotedblright\ with elements $\Lambda =\{\Lambda _{1},\Lambda
_{0}\}$ to distinguish the two states. Central for the following arguments,
here $\Lambda _{1}$ is restricted to be a 1-rank projector, while $\Lambda
_{0}$\ is its complement, $\Lambda _{1}+\Lambda _{0}=\mathrm{I.}$ For
example,%
\begin{equation}
\Lambda _{1}=|\psi _{1}\rangle \langle \psi _{1}|,\ \ \ \ \ \ \ \Lambda _{0}=%
\mathrm{I}-|\psi _{1}\rangle \langle \psi _{1}|=\sum_{i\neq 1}|\psi
_{i}\rangle \langle \psi _{i}|,  \label{Unidimensional}
\end{equation}%
where $\{|\psi _{i}\rangle \}$ is a complete basis.

When the outcome $1$ or $0$ is obtained, Bob guesses the state $\rho _{1}$
or $\rho _{0}$ respectively. Thus, the probability $p_{succ}(\Lambda )$ for
this hypothesis testing scenario is 
\begin{subequations}
\begin{eqnarray}
p_{succ}(\Lambda ) &=&\mathrm{Tr}[\Lambda _{1}\rho _{1}]\frac{1}{2}+\mathrm{%
Tr}[\Lambda _{0}\rho _{0}]\frac{1}{2}, \\
&=&\frac{1}{2}\{1+\mathrm{Tr}[\Lambda _{1}(\rho _{1}-\rho _{0})]\},
\end{eqnarray}%
where we have used that $\Lambda _{1}+\Lambda _{0}=\mathrm{I.}$ Now, we
assume that Bob can choose freely the projectors $\{\Lambda _{1},\Lambda
_{0}\}$ such that $p_{succ}(\Lambda )$ is maximized. The success probability
with respect to all measurements, under the constraint~(\ref{Unidimensional}%
), can then be defined as 
\end{subequations}
\begin{equation}
p_{succ}(\Lambda )\!=\!\frac{1}{2}\max \left\{ 
\begin{array}{c}
1+\max_{\{\Lambda _{1}\}}\mathrm{Tr}[\Lambda _{1}(\rho _{1}-\rho _{0})] \\ 
\\ 
1-\min_{\{\Lambda _{1}\}}\mathrm{Tr}[\Lambda _{1}(\rho _{1}-\rho _{0})]%
\end{array}%
\right. .
\end{equation}%
This expression can be rewritten as%
\begin{equation}
p_{succ}(\Lambda )=\frac{1}{2}(1+\max_{\{\Lambda _{1}\}}|\mathrm{Tr}[\Lambda
_{1}(\rho _{1}-\rho _{0})]|),
\end{equation}%
Using that $\Lambda _{1}$ is a one-dimensional projector [Eq.~(\ref%
{Unidimensional})] and given that the states $\rho $ that maximize $D_{\rho
} $ can always be chosen as pure states [Eq.~(\ref{RhoMax})], it follows that%
\begin{equation}
\max_{\{\Lambda _{1}\}}|\mathrm{Tr}[\Lambda _{1}(\rho _{1}-\rho
_{0})]|=D_{\rho }(\rho _{1},\rho _{0}),  \label{RangoUno}
\end{equation}%
which implies that%
\begin{equation}
p_{succ}(\Lambda )=\frac{1}{2}[1+D_{\rho }(\rho _{1},\rho _{0})].
\label{Testing}
\end{equation}%
Consequently, the proposed distinguishability measure $D_{\rho }$ is related
to the maximum success probability in distinguishing two quantum states in a
quantum hypothesis testing experiment. We notice that when the rank of $%
\Lambda _{1}$ can be greater than one, the success probability $%
p_{succ}(\Lambda ),$ instead of $D_{\rho }(\rho _{1},\rho _{0}),$ is defined
in terms of $D_{\Pi }(\rho _{1},\rho _{0})$ \cite{wildeRep}.

\subsection{Comparison between metrics}

From the previous analysis one can conclude that $D_{\Pi }(\rho _{A},\rho
_{B})$ and $D_{\rho }(\rho _{A},\rho _{B})$ [Eqs.~(\ref{MaxProDef}) and~(\ref%
{DrhoDefinition}) respectively], are intrinsically different
distinguishability measures. Here, we establish when they are equal and how
they differ in general.

Both distinguishability measures always coincide when the Hilbert space
dimension $\dim (\mathcal{H})$ is equal to two and three, 
\begin{equation}
D_{\rho }(\rho _{A},\rho _{B})=D_{\Pi }(\rho _{A},\rho _{B}),\ \ \ \ \ \ \
\dim (\mathcal{H})=2,3.  \label{23}
\end{equation}%
Furthermore, when $\dim (\mathcal{H})\geq 4,$ the inequalities%
\begin{equation}
D_{\rho }(\rho _{A},\rho _{B})\leq D_{\Pi }(\rho _{A},\rho _{B})\leq 
\mathcal{N}D_{\rho }(\rho _{A},\rho _{B})  \label{Constraints}
\end{equation}%
are fulfilled, where the constant $\mathcal{N}$ is%
\begin{equation}
\mathcal{N}=\mathrm{Int}[\dim (\mathcal{H})/2].
\end{equation}%
$\mathrm{Int}[a]$ denotes the integer part of real number $a.$

The conditions under which the equalities in Eq.~(\ref{Constraints}) are
satisfied (higher dimensional spaces, $\dim (\mathcal{H})\geq 4)$ are also
well defined. $D_{\rho }(\rho _{A},\rho _{B})=D_{\Pi }(\rho _{A},\rho _{B})$
when the eigenvalue of $(\rho _{A}-\rho _{B})$ with maximal absolute value
is not degenerate. Equivalently, this occurs when $(\rho _{A}-\rho _{B})$
has a unique positive (or negative) eigenvalue. On the other hand, $D_{\Pi
}(\rho _{A},\rho _{B})=\mathcal{N}D_{\rho }(\rho _{A},\rho _{B})$ when the
eigenvalue of $(\rho _{A}-\rho _{B})$ with maximal absolute value has
degeneracy $\mathcal{N}.$

\textit{Demonstration:} Below we demonstrate the validity of Eqs.~(\ref{23})
and~(\ref{Constraints}). By using Eq.~(\ref{DiferenciaDiagonal}), $(\rho
_{A}-\rho _{B})=\sum_{i}\zeta _{i}\ |i\rangle \langle i|,$ the
projector-based measure [Eq.~(\ref{DPi})], $D_{\Pi }(\rho _{A},\rho
_{B})=(1/2)\mathrm{Tr}|\rho _{A}-\rho _{B}|,$ can be written in terms of the
eigenvalues $\{\zeta _{i}\}$ of $(\rho _{A}-\rho _{B})$\ as%
\begin{equation}
D_{\Pi }(\rho _{A},\rho _{B})=\frac{1}{2}\sum_{i}|\zeta _{i}|=\frac{1}{2}%
\Big{(}\sum_{i=1}^{n_{+}}\zeta _{i}^{(+)}+\sum_{j=1}^{n_{-}}|\zeta
_{j}^{(-)}|\Big{)}.  \label{DProj_Eigen}
\end{equation}%
In the second equality, we split the addition in positive and negative
eigenvalues, $\{\zeta _{i}^{(+)}\}$\ and $\{\zeta _{j}^{(-)}\}$\
respectively. Furthermore, $n_{+}$ and $n_{-}$ count their quantity
respectively, $n_{+}+n_{-}=\dim (\mathcal{H})$ \cite{ceros}. Given that $%
\mathrm{Tr}[(\rho _{A}-\rho _{B})]=0$ it is fulfilled that $%
\sum_{i=1}^{n_{+}}\zeta _{i}^{(+)}=\sum_{j=1}^{n_{-}}|\zeta _{j}^{(-)}|.$
Hence, straightforwardly it follows that%
\begin{equation}
D_{\Pi }(\rho _{A},\rho _{B})=\sum_{i=1}^{n_{+}}\zeta
_{i}^{(+)}=\sum_{j=1}^{n_{-}}|\zeta _{j}^{(-)}|.  \label{DefTraceEigen}
\end{equation}%
On the other hand, Eq.~(\ref{DRhoEigenSol}) tells us that $D_{\rho }(\rho
_{A},\rho _{B})=\max_{\{i\}}\{|\zeta _{i}|\}.$ Consequently, when the number 
$n_{+}$ \textit{or} $n_{-}$ of positive and negative eigenvalues are equal
to \textit{one} both measures coincides, that is,%
\begin{equation}
n_{+}=1\ \ \ \ or\ \ \ \ n_{-}=1\ \ \ \ \Leftrightarrow \ \ \ \ D_{\rho
}=D_{\Pi }.  \label{condition}
\end{equation}%
In fact, in this situation, the unique positive (or negative) eigenvalue,
due to the equality $\sum_{i=1}^{n_{+}}\zeta
_{i}^{(+)}=\sum_{j=1}^{n_{-}}|\zeta _{j}^{(-)}|,$ is also the eigenvalue
with maximal absolute value, which in turn is not degenerate. This condition
can be rephrased as follows: when $(\rho _{A}-\rho _{B})$ has a unique
positive (or negative) eigenvalue, then $D_{\rho }(\rho _{A},\rho
_{B})=D_{\Pi }(\rho _{A},\rho _{B}).$

The previous condition [Eq.~(\ref{condition})] is always fulfilled when $%
\dim (\mathcal{H})=2$ where $n_{+}=n_{-}=1.$ The same occurs when $\dim (%
\mathcal{H})=3$ because it can only occur that $n_{+}=2,$ $n_{-}=1,$ or
complementarily $n_{+}=1,$ $n_{-}=2.$ The same occurs if there is a null
eigenvalue, which implies $n_{+}=1,$ $n_{-}=1.$ Consequently, Eq.~(\ref{23})
is established.

For Hilbert spaces with $\dim (\mathcal{H})\geq 4$ the equality of $D_{\rho
}(\rho _{A},\rho _{B})$ and $D_{\Pi }(\rho _{A},\rho _{B})$ is not valid in
general, but accidentally it occurs when $n_{+}=1$ or\ $n_{-}=1.$ On the
other hand, from Eq.~(\ref{DefTraceEigen}) it follows that $D_{\Pi }(\rho
_{A},\rho _{B})\leq n_{s}\max_{\{i\}}\{|\zeta _{i}|\}=n_{s}D_{\rho }(\rho
_{A},\rho _{B}),$ where $n_{s}$ is the number of positive or negative
eigenvalues and $s=\pm 1$ gives the sign of the eigenvalue with maximal
absolute value. Given that $\sum_{i=1}^{n_{+}}\zeta
_{i}^{(+)}=\sum_{j=1}^{n_{-}}|\zeta _{i}^{(-)}|,$ the maximal possible value
of $n_{s}$ is $\mathcal{N}=\mathrm{Int}[\dim (\mathcal{H})/2]$ \cite{Entero}%
. These results lead to the upper constraint in Eq.~(\ref{Constraints}). It
is achieved when the eigenvalue with maximal absolute value has degeneracy $%
\mathcal{N}.$ Thus, the conditions under which the equalities in Eq.~(\ref%
{Constraints}) are fulfilled are established.

\section{Contractivity under quantum operations}

Here, we characterize the behavior of $D_{\rho }$ under quantum operations.
Since it is based on an infinite norm [see Eq.~(\ref{AlfaLimit})] from Ref.~%
\cite{ruskai} we can anticipate that contractivity is not fulfilled here.
The alternative analysis developed below allows us to establish the states
and maps that lead to maximal violation of contractivity, which further
leads to the formulation of a quantumness measure for non-unital maps.

First, we notice that, given a trace preserving completely positive map, $%
\rho \rightarrow \mathcal{E}(\rho ),$ the projector-based measure [Eqs.~(\ref%
{MaxProDef}) and (\ref{DPi})] always satisfies \textit{contractivity}~\cite%
{nielsen}%
\begin{equation}
D_{\Pi }(\mathcal{E}[\rho _{A}],\mathcal{E}[\rho _{B}])\leq D_{\Pi }(\rho
_{A},\rho _{B}).
\end{equation}%
Hence, the distance between two states can never increase under the action
of the map $\mathcal{E}.$ For the state-based measure [Eqs.~(\ref%
{DrhoDefinition}) and (\ref{DRhoEigenSol})], when the Hilbert space
dimension is $\dim (\mathcal{H})=2$ and $\dim (\mathcal{H})=3,$ we find that%
\begin{equation}
D_{\rho }(\mathcal{E}[\rho _{A}],\mathcal{E}[\rho _{B}])\leq D_{\rho }(\rho
_{A},\rho _{B}).  \label{ContractorRho}
\end{equation}%
This result follows straightforwardly because, with this dimensionality, $%
D_{\rho }(\rho _{A},\rho _{B})=D_{\Pi }(\rho _{A},\rho _{B})$ [Eq.~(\ref{23}%
)]. In addition, contractivity [Eq.~(\ref{ContractorRho})]\ is always
satisfied if $\mathcal{E}$ is a \textit{unital} map, that is, when $\mathcal{%
E}[\mathrm{I}]=\mathrm{I}$ (\textrm{I} is the identity operator). For 
\textit{non-unital} maps, $\mathcal{E}[\mathrm{I}]\neq \mathrm{I},$ and for
higher dimensional spaces $[\dim (\mathcal{H})\geq 4]$, it is possible to
obtain%
\begin{equation}
D_{\rho }(\mathcal{E}[\rho _{A}],\mathcal{E}[\rho _{B}])\leq \mathcal{C}%
D_{\rho }(\rho _{A},\rho _{B}).  \label{InequalDual}
\end{equation}%
The constant $\mathcal{C}$ is bounded as%
\begin{equation}
1<\mathcal{C}\leq \dim (\mathcal{H}),  \label{cota_C}
\end{equation}%
implying that standard contractivity is not fulfilled in general $(\mathcal{C%
}\neq 1).$ Furthermore, $\mathcal{C}$ can be written as%
\begin{equation}
\mathcal{C}=\max_{\{\rho \}}\mathrm{Tr}[V_{\mathcal{E}}\rho
]=\max_{\{k\}}\{v_{k}\}.  \label{C_Eigen}
\end{equation}%
The positive definite operator $V_{\mathcal{E}}$ reads%
\begin{equation}
V_{\mathcal{E}}\equiv \mathcal{E}[\mathrm{I}]=\sum_{\alpha }V_{\alpha
}V_{\alpha }^{\dagger }=\sum_{k}v_{k}|k\rangle \langle k|,
\end{equation}%
where $\{v_{k}\}$ and $\{|k\rangle \}$ are the corresponding eigenvalues and
eigenbasis. Thus, $\mathcal{C}$ is the largest eigenvalue of the operator $%
V_{\mathcal{E}}.$ On the other hand, the set of operators $\{V_{\alpha }\}$
define the Kraus representation~\cite{nielsen} of the map $\mathcal{E}$ and
its dual $\mathcal{E}^{\#}$, the latter being defined by the relation $%
\mathrm{Tr}[A\mathcal{E}[\rho ]]=\mathrm{Tr}[\rho \mathcal{E}^{\#}[A]].$
Explicitly,%
\begin{equation}
\mathcal{E}[\rho ]=\sum_{\alpha }V_{\alpha }\rho V_{\alpha }^{\dagger },\ \
\ \ \ \ \mathcal{E}^{\#}[\rho ]=\sum_{\alpha }V_{\alpha }^{\dagger }\rho
V_{\alpha }.  \label{Kraus}
\end{equation}%
Notice that trace preservation implies $\sum_{\alpha }V_{\alpha }^{\dagger
}V_{\alpha }=\mathrm{I}.$

\textit{Demonstration}: first, we notice that $V_{\mathcal{E}}=\mathcal{E}[%
\mathrm{I}]=\dim (\mathcal{H})\mathcal{E}[\mathrm{I}/\dim (\mathcal{H})].$
Consequently, $\mathrm{Tr}[V_{\mathcal{E}}]=\dim (\mathcal{H}%
)=\sum_{k}v_{k}, $ which, for non-unital maps, supports the inequality Eq.~(%
\ref{cota_C}). Furthermore, considering unital maps, $V_{\mathcal{E}%
}\rightarrow \mathrm{I}$ (which implies $v_{k}=1\ \forall k)$ leading to $%
\mathcal{C}\rightarrow 1.$

Based on the definition~(\ref{DrhoDefinition}) we write%
\begin{eqnarray}
D_{\rho }(\mathcal{E}[\rho _{A}],\mathcal{E}[\rho _{B}])\!\!
&=&\!\max_{\{\rho \}}\left\vert \mathrm{Tr}[\rho (\mathcal{E}[\rho _{A}]-%
\mathcal{E}[\rho _{B}])]\right\vert  \label{Demo} \\
\!\! &=&\!\max_{\{\rho \}}\left\vert \mathrm{Tr}[\rho \mathcal{E}[\rho
_{A}-\rho _{B}]]\right\vert  \notag \\
\!\! &=&\!\max_{\{\rho \}}\left\vert \mathrm{Tr}[\mathcal{E}^{\#}[\rho
](\rho _{A}-\rho _{B})]\right\vert  \notag \\
\!\!\!\! &=&\!\!\max_{\{\rho \}}\Big{(}\mathrm{Tr}[\mathcal{E}^{\#}[\rho ]]|%
\mathrm{Tr}[\rho _{\mathcal{E}}(\rho _{A}-\rho _{B})]|\Big{)},  \notag
\end{eqnarray}%
where we have used that $\mathrm{Tr}[\mathcal{E}^{\#}[\rho ]]>0$ and defined
the state%
\begin{equation}
\rho _{\mathcal{E}}=\rho _{\mathcal{E}}[\rho ]\equiv \frac{\mathcal{E}%
^{\#}[\rho ]}{\mathrm{Tr}[\mathcal{E}^{\#}[\rho ]]}.  \label{RhoE}
\end{equation}%
Given that $\rho _{\mathcal{E}}$ is a positive definite operator with unit
trace, the second factor in the last line of Eq.~(\ref{Demo}) fulfills$\ |%
\mathrm{Tr}[\rho _{\mathcal{E}}(\rho _{A}-\rho _{B})]|\leq D_{\rho }(\rho
_{A},\rho _{B}),$ leading to the inequality%
\begin{equation}
D_{\rho }(\mathcal{E}[\rho _{A}],\mathcal{E}[\rho _{B}])\leq (\max_{\{\rho
\}}\mathrm{Tr}[\mathcal{E}^{\#}[\rho ]])\ D_{\rho }(\rho _{A},\rho _{B}).
\label{Clear}
\end{equation}%
From here, we recover Eq.~(\ref{InequalDual}) with%
\begin{equation}
\mathcal{C}=\max_{\{\rho \}}\mathrm{Tr}[\mathcal{E}^{\#}[\rho ]].
\label{Constant}
\end{equation}%
Using the Kraus representation [Eq.~(\ref{Kraus})], the cyclic property of
the trace operation, and the maximization defined in Appendix~\ref%
{MaxAverage}, this last expression recovers Eq.~(\ref{C_Eigen}).

\subsection{Maximal departure from contractivity}

Given a non-unital map $\mathcal{E}$, the inequality~(\ref{InequalDual})
implies that there may exist (or not) states $\rho _{A}$ and $\rho _{B}$
such that usual contractivity is violated. Taking into account that the
states that maximize the definition of $D_{\rho }$ [Eq.~(\ref{DrhoDefinition}%
)] can always be chosen as pure states [see Eqs.~(\ref{RhoMax}) and (\ref%
{RangoUno})], we expect that contractivity is not fulfilled for states ($%
\rho _{A}$ and $\rho _{B}$) whose \textit{purity} is increased by the map.
Nevertheless, this relation is not valid in general (over the complete set
of possible input states). On the other hand, here we analyze the conditions
under which maximal departure could be achieved, $D_{\rho }(\mathcal{E}[\rho
_{A}],\mathcal{E}[\rho _{B}])=\mathcal{C}D_{\rho }(\rho _{A},\rho _{B}).$

Taking into account the last line of Eq.~(\ref{Demo}), the equality in Eq.~(%
\ref{InequalDual}) is fulfilled when the state $\rho ^{\max }$ that
maximizes the definition of $D_{\rho }(\rho _{A},\rho _{B})$ can be written
as $\rho ^{\max }=\rho _{\mathcal{E}}[\rho _{v}]$ where $\rho _{v}$ is the
state that maximizes Eq.~(\ref{Constant}). The state $\rho _{v}$ can always
be chosen as the projector (or mixed state) associated to the space spanned
by the eigenstate of $V_{\mathcal{E}}$ with maximal eigenvalue [Eq.~(\ref%
{C_Eigen})] (Appendix~\ref{MaxAverage}). Due to the action of $\mathcal{E}%
^{\#}$ [Eq.~(\ref{RhoE})], $\rho _{\mathcal{E}}$ is \textit{in general} a
mixed state. Consequently, maximal departure can be reached under the
following conditions. (i) The eigenvalue of $(\rho _{A}-\rho _{B})$ with
maximal absolute value must be degenerate such that $\rho ^{\max }$ can be
chosen as an arbitrary statistical superposition (mixed state) of the
corresponding eigenvectors (Appendix~\ref{MaxAverage}). (ii) The equality $%
\rho ^{\max }=\rho _{\mathcal{E}}[\rho _{v}]$ must be fulfilled.

\subsection{Witnessing maximal violation of contractivity\label{NoUnico}}

For an arbitrary map $\mathcal{E}$ the previous conditions could not be
fulfilled. In such a case, maximal violation of contractivity is not
observed. In contrast, here we demonstrate that by adding a passive
ancillary system, maximal departure from contractivity is always achieved.

For simplicity, the ancilla is taken as a two-level system with associated
basis of states $\{|\pm \rangle \}.$ The map is extended to the
\textquotedblleft system-ancilla\textquotedblright\ Hilbert space as%
\begin{equation}
\mathcal{\tilde{E}}=\mathcal{E}\otimes \mathrm{I}_{a},  \label{ExtendedMap}
\end{equation}%
where $\mathrm{I}_{a}$ is the identity operator for the ancilla system.
Furthermore, we consider the states%
\begin{equation}
\rho _{A}=\frac{\mathrm{I}}{\dim (\mathcal{H})}\otimes |+\rangle \langle
+|,\ \ \ \ \ \rho _{B}=\frac{\mathrm{I}}{\dim (\mathcal{H})}\otimes
|-\rangle \langle -|.  \label{IAB}
\end{equation}%
Therefore, it is simple to obtain%
\begin{equation}
\mathcal{\tilde{E}}[\rho _{A}]-\mathcal{\tilde{E}}[\rho _{B}]=\frac{1}{\dim (%
\mathcal{H})}V_{\mathcal{E}}\otimes (|+\rangle \langle +|-|-\rangle \langle
-|),
\end{equation}%
where we have used that $\mathcal{E}[\mathrm{I}]=V_{\mathcal{E}}.$ From the
previous two expressions, it follows that $D_{\rho }(\rho _{A},\rho
_{B})=1/\dim (\mathcal{H})$ and $D_{\rho }(\mathcal{\tilde{E}}[\rho _{A}],%
\mathcal{\tilde{E}}[\rho _{B}])=\mathcal{C}/\dim (\mathcal{H})$ where $%
\mathcal{C}$ is defined by Eq.~(\ref{C_Eigen}). Consequently,%
\begin{equation}
D_{\rho }(\mathcal{\tilde{E}}[\rho _{A}],\mathcal{\tilde{E}}[\rho _{B}])=%
\mathcal{C}D_{\rho }(\rho _{A},\rho _{B}).  \label{EqualMax}
\end{equation}%
Consistently, the conditions (i) and (ii) previously defined are satisfied.
Furthermore, in agreement with the qualitative argument based on the purity
of the states, here $\mathrm{Tr}[(\mathcal{\tilde{E}}[\rho _{A}])^{2}]+%
\mathrm{Tr}[(\mathcal{\tilde{E}}[\rho _{B}])^{2}]>\mathrm{Tr}[\rho _{A}^{2}]+%
\mathrm{Tr}[\rho _{B}^{2}].$ On the other hand, for the same states [Eq.~(%
\ref{IAB})] we have $D_{\Pi }(\rho _{A},\rho _{B})=1$ and $D_{\Pi }(\mathcal{%
\tilde{E}}[\rho _{A}],\mathcal{\tilde{E}}[\rho _{B}])=\mathrm{Tr}[V_{%
\mathcal{E}}]/\dim (\mathcal{H})=1.$

It is important to notice that the states $\rho _{A}$ and $\rho _{B}$ that
lead to the previous result are not unique. In fact, under the replacements $%
\rho _{A}\rightarrow (1-w)\varrho _{sa}+w\rho _{A}$ and $\rho
_{B}\rightarrow (1-w)\varrho _{sa}+w\rho _{B},$ where $0<w\leq 1$ and $%
\varrho _{sa}$ is an arbitrary system-ancilla state, one again arrives at
the equality~(\ref{EqualMax}).

\subsection{Quantumness of non-unital maps}

The classicality, or complementarily the quantumness, of a given open system
evolution can be tackled from different perspectives~\cite%
{nori,franco,ChenChen,fata,lika,Szanko,sun,Liu,maps,Dq}. Consistent with
Refs.~\cite{nori,Dq}, here a map $\rho \rightarrow \mathcal{E}(\rho )$ with
the structure%
\begin{equation}
\mathcal{E}(\rho )=\sum_{c}p_{c}U_{c}\rho U_{c}^{\dag },  \label{clasico}
\end{equation}%
where $U_{c}$\ is a unitary transformation and whose weigh is $p_{c},$\ is
read as a classical one. In fact, this structure can always be implemented
without involving any quantum feature of the environment. Notice that all
maps that admit this classical interpretation are also unital (the inverse
implication in general is not true, see for example~\cite{clerck}).
Consequently, in contrast with $D_{\Pi },$ the lack of contractivity of $%
D_{\rho }$ witnesses the non-classicality of non-unital maps. This property
allows us to introduce a degree of map quantumness $\mathcal{M}_{Q},$ which
gives one the main supports of the present approach. Given that the constant 
$\mathcal{C}$\ measures the maximal departure from contractivity, $\mathcal{M%
}_{Q}$ is defined as%
\begin{equation}
\mathcal{M}_{Q}\equiv \mathcal{C}-1=\max_{\{\rho \}}|\mathrm{Tr}[\mathcal{E}%
^{\#}[\rho ]]-1|,  \label{Mq}
\end{equation}%
where the equality is based on Eq.~(\ref{Constant}). Furthermore, it is
bounded as $0\leq \mathcal{M}_{Q}\leq \dim (\mathcal{H})-1.$

Using the relation between a map and its dual, Eq.~(\ref{Mq}) can
equivalently be rewritten as%
\begin{equation}
\frac{\mathcal{M}_{Q}}{\dim (\mathcal{H})}=D_{\rho }(\mathcal{E}[\rho _{%
\mathrm{I}}],\rho _{\mathrm{I}}),  \label{Mq_Dinfinita}
\end{equation}%
where $\rho _{\mathrm{I}}\equiv \mathrm{I}/\dim (\mathcal{H})$ is the
maximal mixed state. This equality explicitly shows the role of $D_{\rho }$
in the present definition. Furthermore, it allows to understand the scheme
that permits its determination [Eq.~(\ref{EqualMax})]. In fact, the states~(%
\ref{IAB}) involve the\ (system) maximally mixed state. They lead to maximal
departure from contractivity but, in addition, they lead to $\mathcal{M}%
_{Q}=0$ [Eq.~(\ref{Mq_Dinfinita})] when the map is unital.

Even when $D_{\rho }$ is contractive when $\dim (\mathcal{H})=2$ and $\dim (%
\mathcal{H})=3,$ $\mathcal{M}_{Q}$ can be determine in these cases because
the extra ancilla leads to a higher dimensional space (see Sec.~\ref%
{Thermatal}). With this dimensionality, the constant $\mathcal{C}$ must be
read from the general expression~(\ref{C_Eigen}). On the other hand, we
remark that $\mathcal{M}_{Q}$ also applies to time-dependent open system
dynamics after identifying the map $\mathcal{E}$ with the propagator of the
system density matrix. Eq.~(\ref{Mq_Dinfinita}) also recovers the degree of
environment quantumness introduced in Ref.~\cite{Dq} when studying
continuous-in-time evolutions characterized by a unique stationary state
[see analysis below Eq.~(\ref{Dq})].

\section{Examples}

Here we characterize the proposed distinguishability measure for some
particular quantum states. In addition, its behavior under different
completely positive maps is studied in detail.

\subsection{Particular cases}

$\blacksquare $ When both states are pure, $\rho _{A}=|\psi _{A}\rangle
\langle \psi _{A}|,\ \rho _{B}=|\psi _{B}\rangle \langle \psi _{B}|,$ from
Eq.~(\ref{DrhoDefinition}) we get%
\begin{equation}
D_{\rho }(\rho _{A},\rho _{B})=\max_{\{\rho \}}\left\vert \langle \psi
_{A}|\rho |\psi _{A}\rangle -\langle \psi _{B}|\rho |\psi _{B}\rangle
\right\vert .
\end{equation}%
This expression can be solved after calculating the eigenvalues $\zeta $\
defined by $(\rho _{A}-\rho _{B})|\psi \rangle =\zeta |\psi \rangle ,$ where 
$|\psi \rangle =a|\psi _{A}\rangle +b|\psi _{B}\rangle .$ We get $\zeta =\pm 
\sqrt{1-|\langle \psi _{A}|\psi _{B}\rangle |^{2}}.$ The rest of the
eigenvalues, $\zeta =0,$ correspond to eigenvectors that are perpendicular
to the plane spanned by $|\psi _{A}\rangle $ and $|\psi _{B}\rangle .$ Thus,
from Eq.~(\ref{DRhoEigenSol}) it follows%
\begin{equation}
D_{\rho }(\rho _{A},\rho _{B})=\sqrt{1-|\langle \psi _{A}|\psi _{B}\rangle
|^{2}}.  \label{Fidelity}
\end{equation}%
Given that $D_{\Pi }(\rho _{A},\rho _{B})=\sqrt{1-|\langle \psi _{A}|\psi
_{B}\rangle |^{2}}$ \cite{nielsen}, $D_{\rho }(\rho _{A},\rho _{B})=D_{\Pi
}(\rho _{A},\rho _{B}).$ In fact, $(\rho _{A}-\rho _{B})$ has a unique
positive (negative) eigenvalue [see Eq.~(\ref{condition})].

For orthogonal states, Eq.~(\ref{Fidelity}) leads to%
\begin{equation}
\langle \psi _{A}|\psi _{B}\rangle =0,\ \ \ \ \Rightarrow \ \ \ \ D_{\rho
}(\rho _{A},\rho _{B})=1.
\end{equation}%
Nevertheless, the inverse implication is not valid, that is, $D_{\rho }(\rho
_{A},\rho _{B})=1$ does not imply that $\rho _{A}$ and $\rho _{B}$ are pure
states. Take for example $\rho _{A}=|\psi _{A}\rangle \langle \psi _{A}|$ and%
$\ \rho _{B}=\sum_{k}w_{k}|\psi _{B}^{k}\rangle \langle \psi _{B}^{k}|$
where the positive weights are normalized, $\sum_{k}w_{k}=1,$ and $\langle
\psi _{B}^{k}|\psi _{A}\rangle =0\ \forall k.$

In general, it is simple to realize that $D_{\rho }(\rho _{A},\rho _{B})=1$
if and only if $\rho _{A}$ and $\rho _{B}$ have support on orthogonal
subspaces and $\rho _{A}$ \textit{or} $\rho _{B}$ is a pure state. Instead, $%
D_{\Pi }(\rho _{A},\rho _{B})=1$, whenever $\rho _{A}$ and $\rho _{B}$ have
support on orthogonal subspaces.

$\blacksquare $ Here we consider two qubit states,%
\begin{equation}
\rho _{A}=(1/2)(\mathrm{I}+\mathbf{\alpha }\cdot \mathbf{\sigma }),\ \ \ \ \
\rho _{B}=(1/2)(\mathrm{I}+\mathbf{\beta }\cdot \mathbf{\sigma }),
\end{equation}%
where $\mathbf{\alpha }$ and $\mathbf{\beta }$\ are the Bloch vectors and $%
\mathbf{\sigma }$ is the vector of Pauli matrices. Then, $\rho _{A}-\rho
_{B}=(1/2)(\mathbf{\alpha -\beta })\cdot \mathbf{\sigma =}(1/2)|\mathbf{%
\alpha -\beta }|(\mathbf{n}\cdot \mathbf{\sigma )},$ where $\mathbf{n=}(%
\mathbf{\alpha -\beta })/|\mathbf{\alpha -\beta }|.$ Given that the
eigenvalues of $(\mathbf{n}\cdot \mathbf{\sigma )}$ are $\pm 1,$ it follows%
\begin{equation}
D_{\rho }(\rho _{A},\rho _{B})=\frac{1}{2}|\mathbf{\alpha -\beta }|.
\label{DrhoQubits}
\end{equation}%
In an alternative way, this result explicitly confirms that when $\dim (%
\mathcal{H})=2$ both measures coincides: in fact, $D_{\rho }(\rho _{A},\rho
_{B})=D_{\Pi }(\rho _{A},\rho _{B})=(1/2)|\mathbf{\alpha -\beta }|$ \cite%
{nielsen}.

$\blacksquare $ Now we consider that one of the density matrices is the
maximally mixed state. Under the replacements $\rho _{A}\rightarrow \varrho
, $ where $\varrho $ is an arbitrary density matrix, and $\rho
_{B}\rightarrow \rho _{\mathrm{I}}=\mathrm{I}/\dim (\mathcal{H}),$ from Eq.~(%
\ref{DRhoEigenSol}), we get%
\begin{equation}
D_{\rho }(\varrho ,\rho _{\mathrm{I}})=\max_{\{i\}}\left\{ \left\vert
\lambda _{i}-\frac{1}{\dim (\mathcal{H})}\right\vert \right\} ,
\end{equation}%
where $\{\lambda _{i}\}$ are the eigenvalues of $\varrho .$ This expression
can be rewritten as%
\begin{equation}
D_{\rho }(\varrho ,\rho _{\mathrm{I}})=\frac{1}{\dim (\mathcal{H})}\max (%
\mathcal{D}_{\rho }^{\max },\mathcal{D}_{\rho }^{\min }),  \label{Dq}
\end{equation}%
where the coefficients are 
\begin{subequations}
\begin{eqnarray}
\mathcal{D}_{\rho }^{\max } &\equiv &\dim (\mathcal{H})\max \{\lambda
_{i}\}-1, \\
\mathcal{D}_{\rho }^{\min } &\equiv &1-\dim (\mathcal{H})\min \{\lambda
_{i}\}.
\end{eqnarray}%
Here, $\max \{\lambda _{i}\}$\ and $\min \{\lambda _{i}\}$ are the maximal
and minimal eigenvalues of $\varrho .$ These expressions recover the degree
of environment quantumness $D_{Q}$ introduced in Ref.~\cite{Dq}. With the
present notation it can be written as $D_{Q}=\dim (\mathcal{H})D_{\rho }(%
\tilde{\rho}_{\infty },\rho _{\mathrm{I}}),$ where $\tilde{\rho}_{\infty
}=\lim_{t\rightarrow \infty }\rho _{t}$ (disregarding a technical
time-inversion operation) is the system stationary state. Under the
identification $\tilde{\rho}_{\infty }\rightarrow \mathcal{E}[\rho _{\mathrm{%
I}}],$ this last expression for $D_{Q}$\ assumes the structure of Eq.~(\ref%
{Mq_Dinfinita}).

$\blacksquare $ Take both density matrixes as diagonal ones, with 
\end{subequations}
\begin{eqnarray*}
\rho _{A} &=&(1/10)\mathrm{diag}\{5,2,2,1\}, \\
\rho _{B} &=&(1/4)\mathrm{diag}\{1,1,1,1\}.
\end{eqnarray*}%
Given that $\rho _{B}$ is the maximally mixed state, $\rho _{A}$ can be read
as an arbitrary quantum state written in its eigenbasis. We notice that $%
(\rho _{A}-\rho _{B})$ only has one positive eigenvalue. This eigenvalue is
not degenerate and coincides with the eigenvalue with maximal absolute
value. Thus, both measures [Eqs.~(\ref{DPi}) and~(\ref{DRhoEigenSol})]
coincide. In fact,%
\begin{equation}
D_{\rho }(\rho _{A},\rho _{B})=D_{\Pi }(\rho _{A},\rho _{B})=0.25.
\end{equation}

$\blacksquare $ Instead, taking%
\begin{eqnarray*}
\rho _{A} &=&(1/10)\mathrm{diag}\{5,3,1,1\}, \\
\rho _{B} &=&(1/4)\mathrm{diag}\{1,1,1,1\},
\end{eqnarray*}%
it follows that $\rho _{A}-\rho _{B}$ has two positive and two negative
eigenvalues. In this case, both measures differ [Eqs.~(\ref{DPi}) and~(\ref%
{DRhoEigenSol})]. We get%
\begin{equation}
D_{\rho }(\rho _{A},\rho _{B})=0.25<D_{\Pi }(\rho _{A},\rho _{B})=0.3.
\end{equation}

$\blacksquare $ In this example%
\begin{eqnarray*}
\rho _{A} &=&(1/10)\mathrm{diag}\{4,4,1,1\}, \\
\rho _{B} &=&(1/4)\mathrm{diag}\{1,1,1,1\}.
\end{eqnarray*}%
Hence, $\rho _{A}-\rho _{B}$ has two degenerate positive eigenvalues, as
well as two degenerate negative eigenvalues. In this case, both measures
differ [Eqs.~(\ref{DPi}) and~(\ref{DRhoEigenSol})]. It is fulfilled that $%
0.15=D_{\rho }(\rho _{A},\rho _{B})<D_{\Pi }(\rho _{A},\rho _{B})=0.3.$ In
addition, the eigenvalue with maximal absolute value has degeneracy equal to
two. Consistently with Eq.~(\ref{Constraints}) it is fulfilled that%
\begin{equation}
D_{\Pi }(\rho _{A},\rho _{B})=0.3=2D_{\rho }(\rho _{A},\rho _{B}).
\end{equation}

$\blacksquare $ Here we take the quantum states 
\begin{subequations}
\label{QuantumStates}
\begin{eqnarray}
\rho _{A} &=&\frac{1}{2}(\mathrm{I}_{2}+r\sigma _{z})\otimes \frac{1}{2}(%
\mathrm{I}_{2}+r\sigma _{z}), \\
\rho _{B} &=&\frac{1}{4}(\mathrm{I}_{4}+s\sigma _{x}\otimes \sigma _{x}),
\end{eqnarray}%
where the parameters are constrained as $0\leq r\leq 1$ and $0\leq s\leq 1.$
The dimensionality of the identity matrix $\mathrm{I}$ is denoted with its
subindex. Furthermore, $\sigma _{i}$ are the Pauli matrices. We notice that
while $\rho _{A}$ (a separable state) is diagonal in the natural basis, $%
\rho _{B}$ is diagonal in the Bell basis. The four eigenvalues of $(\rho
_{A}-\rho _{B})$ are $\{\zeta _{i}\}=(1/4)\{(\pm s-r^{2}),(r^{2}\pm \sqrt{%
4r^{2}+s^{2}})\}.$ Hence, $D_{\rho }(\rho _{A},\rho _{B})$ and $D_{\Pi
}(\rho _{A},\rho _{B})$ follow from Eqs.~(\ref{DRhoEigenSol}) and~(\ref%
{DefTraceEigen}) respectively. After some algebra we find that $D_{\rho
}(\rho _{A},\rho _{B})=D_{\Pi }(\rho _{A},\rho _{B})$ if $s\leq r^{2}.$ In
Fig.~1 we plot both distinguishability measures as a function of $s$ for two
different values of $r.$ Consistently, the behaviors confirm both the
inequalities $(\mathcal{N}=2)$ and equalities defined by Eq.~(\ref%
{Constraints}). 
\begin{figure}[t]
\includegraphics[bb=47 872 726
1140,angle=0,width=8.5cm]{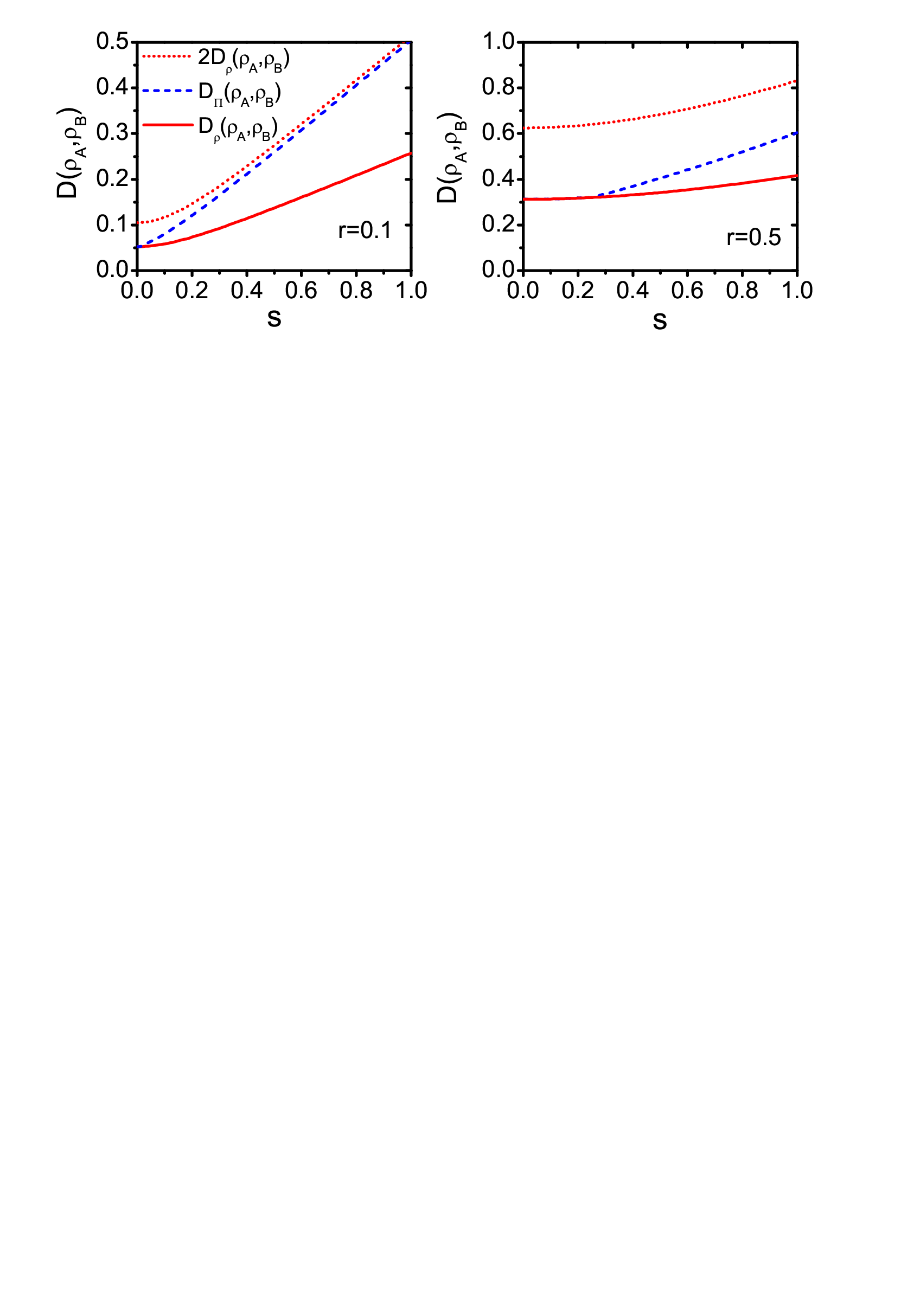}
\caption{Distances between the quantum states defined by Eq.~(\protect\ref%
{QuantumStates}). The full lines correspond to $D_{\protect\rho }(\protect%
\rho _{A},\protect\rho _{B}),$ the dashed lines to $D_{\Pi }(\protect\rho %
_{A},\protect\rho _{B}),$ while the dotted lines correspond to $2D_{\protect%
\rho }(\protect\rho _{A},\protect\rho _{B}).$ The figures show the
dependence with the parameter $s$ associated to $\protect\rho _{B}$. The
left and right panels correspond to $r=0.1$ and $r=0.5$ respectively, where $%
r$ is the parameter associated to $\protect\rho _{A}.$}
\end{figure}

\subsection{Depolarizing maps}

Depolarizing maps (in any Hilbert space dimension) can be defined as 
\end{subequations}
\begin{equation}
\rho \rightarrow \mathcal{E}_{w}[\rho ]=w\rho +(1-w)\frac{\mathrm{I}}{\dim (%
\mathcal{H})},
\end{equation}%
where $0\leq w<1.$ Given that this map is unital~\cite{nielsen}, our
previous analysis guarantees that contractivity is fulfilled [Eq.~(\ref%
{ContractorRho})]. In fact, by writing $\rho _{A}-\rho _{B}=\sum \xi
_{i}|i\rangle \langle i|$ it follows that $\mathcal{E}[\rho _{A}]-\mathcal{E}%
[\rho _{B}]=w(\rho _{A}-\rho _{B})=w\sum \xi _{i}|i\rangle \langle i|.$
Given that $w|\xi _{i}|<|\xi _{i}|$ $\forall i,$ using Eq.~(\ref%
{DRhoEigenSol}), it follows that%
\begin{equation}
D_{\rho }(\mathcal{E}_{w}(\rho _{A}),\mathcal{E}_{w}(\rho _{B}))<D_{\rho
}(\rho _{A},\rho _{B}),
\end{equation}%
where $D_{\rho }(\mathcal{E}_{w}(\rho _{A}),\mathcal{E}_{w}(\rho
_{B}))=w\max_{\{i\}}\{|\zeta _{i}|\}$ while $D_{\rho }(\rho _{A},\rho
_{B})=\max_{\{i\}}\{|\zeta _{i}|\}.$

\subsection{Zero temperature qubit map\label{Thermatal}}

A qubit system coupled to a zero temperature reservoir can be described by
the map $\mathcal{E}[\rho ]=V_{0}\rho V_{0}^{\dagger }+V_{1}\rho
V_{1}^{\dagger },$ with Kraus operators%
\begin{equation}
V_{0}=\left( 
\begin{array}{cc}
\sqrt{1-\gamma } & 0 \\ 
0 & 1%
\end{array}%
\right) ,\ \ \ \ \ \ V_{1}=\left( 
\begin{array}{cc}
0 & 0 \\ 
\sqrt{\gamma } & 0%
\end{array}%
\right) ,  \label{Kraus2}
\end{equation}%
where $\gamma \in \lbrack 0,1].$ The action over an arbitrary state $\rho $
is%
\begin{equation}
\rho =\left( 
\begin{array}{cc}
p & c \\ 
c^{\ast } & q%
\end{array}%
\right) \ \ \rightarrow \ \ \mathcal{E}[\rho ]=\left( 
\begin{array}{cc}
(1-\gamma )p & \sqrt{1-\gamma }c \\ 
\sqrt{1-\gamma }c^{\ast } & q+\gamma p%
\end{array}%
\right) ,  \label{ThermalMap}
\end{equation}%
where $p$ and $q$ denote populations while $c$ denotes coherence. Notice
that the parameter $\gamma $ gives the probability for a transition from the
upper to the lower level, $|+\rangle \rightarrow |-\rangle .$

Consistent with the trace preservation property, it is fulfilled that $%
V_{0}^{\dagger }V_{0}+V_{1}^{\dagger }V_{1}=\mathrm{I.}$ On the other hand,%
\begin{equation}
\mathcal{E}[\mathrm{I}]=V_{0}V_{0}^{\dagger }+V_{1}V_{1}^{\dagger }=\left( 
\begin{array}{cc}
1-\gamma & 0 \\ 
0 & 1+\gamma%
\end{array}%
\right) \neq \mathrm{I}.  \label{V2}
\end{equation}%
Thus, the map is not unital (also non-classical). Nevertheless, given the
system dimensionality, $\dim (\mathcal{H})=2,$ contractivity must be
fulfilled [Eq.~(\ref{ContractorRho})]. This property is corroborated in
Appendix~\ref{ThermalD2}.

Here we study the two qubits map%
\begin{equation}
\mathcal{E}=\mathcal{E}_{a}\otimes \mathcal{E}_{b},  \label{Ebipartite}
\end{equation}%
where the maps $\mathcal{E}_{a}$ and $\mathcal{E}_{b}$ are defined by the
Kraus operators~(\ref{Kraus2}) under the replacements $\gamma \rightarrow
\gamma _{a}$ and $\gamma \rightarrow \gamma _{b}$ respectively.

Instead of proposing a set of states $\rho _{A}$ and $\rho _{B},$ we write
their difference $(\rho _{A}-\rho _{B})$ in its proper eigenbasis as%
\begin{equation}
\Delta =\rho _{A}-\rho _{B}=\left( 
\begin{array}{cccc}
x & 0 & 0 & 0 \\ 
0 & y & 0 & 0 \\ 
0 & 0 & z & 0 \\ 
0 & 0 & 0 & -(x+y+z)%
\end{array}%
\right) .  \label{Delta}
\end{equation}%
Under appropriate constraints on these parameters [eigenvalues $x,$ $y,$ $z$
and $-(x+y+z)],$ 
\begin{subequations}
\label{Domain}
\begin{equation}
|x|\leq 1,\ \ \ \ |y|\leq 1,\ \ \ \ |z|\leq 1,\ \ \ \ |x+y+z|\leq 1,
\end{equation}%
jointly with 
\begin{equation}
|x+y|\leq 1,\ \ \ \ |x+z|\leq 1,\ \ \ \ |y+z|\leq 1,
\end{equation}
the matrix $\Delta $ represents a difference of two arbitrary density
matrices [see derivation in Appendix~\ref{Space}]. From its definition~(\ref%
{DRhoEigenSol}), the distance between the input states $[D_{\rho }(\rho
_{A},\rho _{B})=D_{\rho }(\Delta )]$ is 
\end{subequations}
\begin{equation}
D_{\rho }(\Delta )=\max \{|x|,|y|,|z|,|x+y+z|\}.  \label{DIn4}
\end{equation}
\begin{figure}[t]
\includegraphics[bb=0 0 380 290,angle=0,width=6cm]{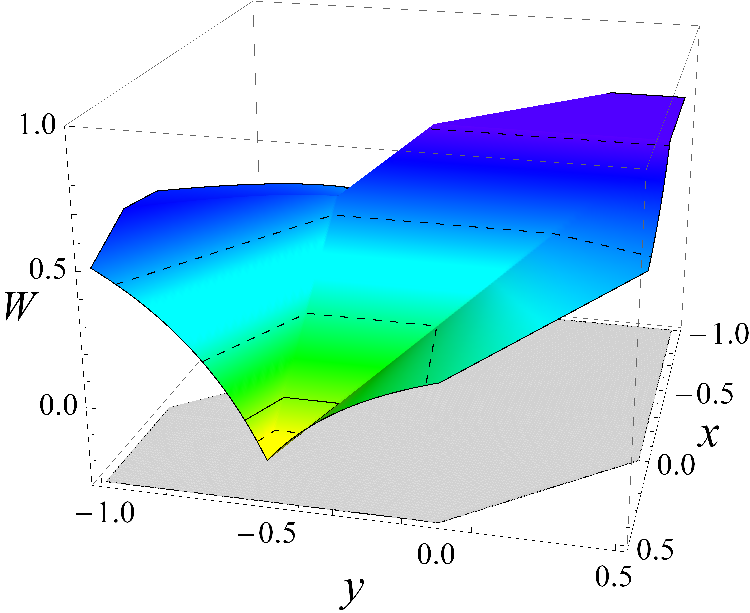}
\caption{Witness $W$ [Eq.~(\protect\ref{W})] for the two qubit map~(\protect
\ref{Ebipartite}) as a function of $(x,y)$ and fixed $z$ [Eq.~(\protect\ref%
{Delta})]. The map parameters are $\protect\gamma _{a}=1/2$ and $\protect%
\gamma _{b}=1/4.$ The horizontal full line corresponds to the level curve $%
W=0.$ The gray plane corresponds to the domain of $(x,y)$ given that here $%
z=0.5.$ }
\label{Landscape1}
\end{figure}

In order to solve the action of the map on the difference of states $\Delta $
we need to specify explicitly the basis where it is diagonal. For
simplicity, we take the same basis where the Kraus operator are defined, $%
\{|++\rangle ,|+-\rangle ,|-+\rangle ,|--\rangle \}.$ In this case, the
application of the map~(\ref{Ebipartite}) over $\Delta ,$ leads to a
diagonal matrix $\mathcal{E}[\Delta ]$ whose four elements are 
\begin{subequations}
\begin{eqnarray}
\mathcal{E}[\Delta ]_{++} &=&(1-\gamma _{a})(1-\gamma _{b})x, \\
\mathcal{E}[\Delta ]_{+-} &=&(1-\gamma _{a})(x\gamma _{b}+y), \\
\mathcal{E}[\Delta ]_{-+} &=&(1-\gamma _{b})(x\gamma _{a}+z), \\
\mathcal{E}[\Delta ]_{--} &=&-(1-\gamma _{a}\gamma _{b})x-(1-\gamma
_{a})y-(1-\gamma _{b})z.\ \ \ \ \ \ \ \ \ 
\end{eqnarray}%
We notice that here the symmetry under interchange of subsystems, $%
a\leftrightarrow b,$ is consistently fulfilled under the parameter changes $%
\gamma _{a}\leftrightarrow \gamma _{b}$ and $y\leftrightarrow z.$ The
distance between the output states, from~(\ref{DRhoEigenSol}), can be
written as 
\end{subequations}
\begin{equation}
D_{\rho }(\mathcal{E}[\Delta ])=\max_{\{s,s^{\prime }\}}\{|\mathcal{E}%
[\Delta ]_{ss^{\prime }}|\},\ \ \ \ s=\pm 1,\ \ \ s^{^{\prime }}=\pm 1.
\label{DOut4}
\end{equation}

Both $D_{\rho }(\Delta )$ and $D_{\rho }(\mathcal{E}[\Delta ])$ [Eqs.~(\ref%
{DIn4}) and~(\ref{DOut4})] depend on $(x,y,z).$ This dependence labels
different possible states $\rho _{A}$ and $\rho _{B}.$ In order to quantify
the violation of (standard) contractivity [Eq.~(\ref{InequalDual})] we
introduce the (dimensionless) witness%
\begin{equation}
W\equiv \frac{D_{\rho }(\rho _{A},\rho _{B})-D_{\rho }(\mathcal{E}[\rho
_{A}],\mathcal{E}[\rho _{B}])}{D_{\rho }(\rho _{A},\rho _{B})(\mathcal{C}-1)}%
.  \label{W}
\end{equation}%
If $W\geq 0$ usual contractivity is fulfilled. Whenever $W<0$ usual
contractivity is not fulfilled. When $W=-1$ the maximal violation of
contractivity is achieved. In fact, in this case $D_{\rho }(\mathcal{E}[\rho
_{A}],\mathcal{E}[\rho _{B}])=\mathcal{C}D_{\rho }(\rho _{A},\rho _{B}).$
Furthermore, notice that $W=W(x,y,z)$ where $D_{\rho }(\rho _{A},\rho
_{B})=D_{\rho }(\Delta )$ [Eq.~(\ref{DIn4})] and $D_{\rho }(\mathcal{E}[\rho
_{A}],\mathcal{E}[\rho _{B}])=D_{\rho }(\mathcal{E}[\Delta ])$ [Eq.~(\ref%
{DOut4})].

For the bipartite map~(\ref{Ebipartite}) the constant $\mathcal{C},$ from
Eqs.~(\ref{C_Eigen}) and~(\ref{V2}),\ is%
\begin{equation}
\mathcal{C}=(1+\gamma _{a})(1+\gamma _{b})\leq 4.
\end{equation}%
Consistent with our definitions [$\mathcal{M}_{Q}=\mathcal{C}-1,$ see Eq.~(%
\ref{Mq})], classicality is only achieved when $\gamma _{a}=\gamma _{b}=0,$
which reduces the map [Eq.~(\ref{Ebipartite})] to the identity.

\begin{figure}[t]
\includegraphics[bb=0 0 380 310,angle=0,width=6cm]{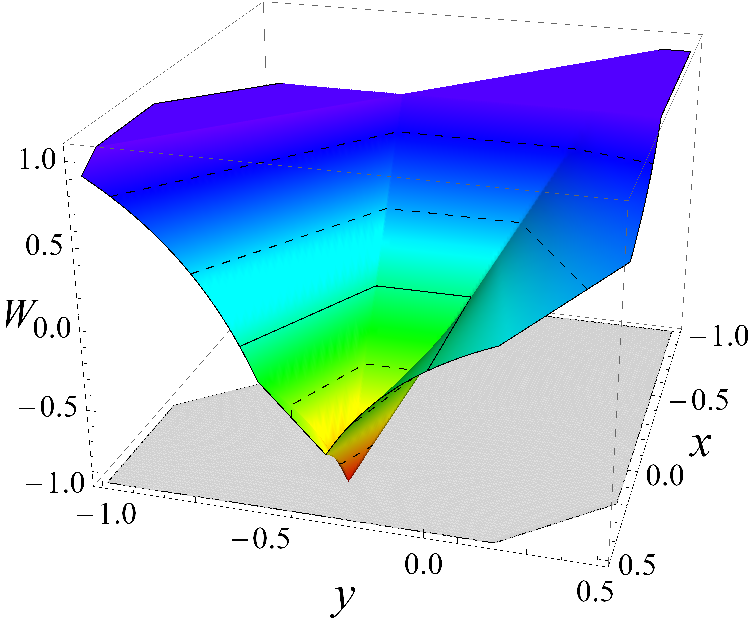}
\caption{Witness $W$ [Eq.~(\protect\ref{W})] for the two qubit map~(\protect
\ref{Ebipartite}) as a function of $(x,y)$ and fixed $z$ [Eq.~(\protect\ref%
{Delta})]. The map parameters are $\protect\gamma _{a}=1/2$ and $\protect%
\gamma _{b}=0.$ The horizontal full line corresponds to the level curve $%
W=0. $ The gray plane corresponds to the domain of $(x,y)$ given that here $%
z=0.3. $ }
\label{Landscape2}
\end{figure}

In Figs.~\ref{Landscape1} and ~\ref{Landscape2} we plot the contractivity
witness $W$ as a function of $(x,y)$ and fixed $z.$ Given $z,$ the domain of
the $(x,y)$ variables corresponds to the surface defined by $z=$constant in
the three dimensional body defined by Eq. (\ref{Domain}) (see Fig.~\ref%
{Solid3D} in Appendix~\ref{Space}). 

In Fig.~\ref{Landscape1} the map parameters [Eqs.~(\ref{ThermalMap}) and~(%
\ref{Ebipartite})] are $\gamma _{a}=1/2$ and $\gamma _{b}=1/4.$ Furthermore,
we take $z=0.5.$ Depending on the values of $(x,y)$ we observe a transition
between contractivity $(W>0)$ and its violation $(W<0).$ Furthermore, we
observe that the limit of maximal departure from contractivity is not
achieved $(W\neq -1).$ We checked that these properties remain the same when
considering other possible values of $z.$

In general, the dependence of $W$ on $(x,y,z)$ defines a complex landscape.
It may include regions where $W$ is constant or even develops non-smooth
non-derivable behaviors. These features are inherited from the expressions
for the input and output distances, Eqs.~(\ref{DIn4}) and~(\ref{DOut4}),
which involve a maximization associated to the definition of $D_{\rho }.$
Given this feature, in general it is not easy or even possible to infer
(analytically) general properties of $W$ as a function of the underlying
parameters, here $\gamma _{a}$\ and $\gamma _{b}.$ Nevertheless, for this
example it is possible to check the following properties.

Assuming that $\gamma _{b}\leq \gamma _{a},$ the witness $W=W(x,y,z)$
assumes its minimal value, 
\begin{equation}
W_{\min }=1-\frac{2\gamma _{a}}{(\gamma _{a}+\gamma _{b}+\gamma _{a}\gamma
_{b})},  \label{Wmin}
\end{equation}%
when $x=z,$ $y=-z$ and \textit{arbitrary}$\ z$ in its domain. For this
choice, its domain is $|z|\leq 1/2$ [see Eq.~(\ref{Domain})]. From the
expression of $W_{\min }$ it follows that when%
\begin{equation}
\gamma _{b}<\frac{\gamma _{a}}{(1+\gamma _{a})},
\end{equation}%
there exist input states [values $(x,y,z)=(z,-z,z),\ $with $z\neq 0$] where
contractivity is not fulfilled $(W_{\min }<0).$ The parameters of Fig.~\ref%
{Landscape1} are in this regime, where $W_{\min }\cong -0.14$ at $x=0.5,$ $%
y=-0.5,$ $z=0.5.$

From Eq.~(\ref{Wmin}) it follows that maximal departure $(W_{\min }=-1)$ can
only be achieved when $\gamma _{b}=0.$ Hence, the subsystem $b$ can be read
as the passive ancillary system associated to the scheme of Sec.~\ref%
{NoUnico}, which allows to determine the quantumness of the two-dimensional
qubit map [Eq.~(\ref{ThermalMap})],%
\begin{equation}
\mathcal{C}=(1+\gamma _{a}),\ \ \ \ \ \ \ \ \mathcal{M}_{Q}=\gamma _{a}.
\end{equation}%
In Fig.~\ref{Landscape2} we check this regime. The map parameters are $%
\gamma _{a}=1/2,$ $\gamma _{b}=0.$ Furthermore, $z=0.3.$ Consistently, when $%
x=0.3,$ $y=-0.3,$ it is achieved $W=-1.$ Here, the degeneracy of the value
of $z$ for getting $W=-1$ $[(x,y,z)=(z,-z,z)]$ can straightforwardly be
related to the non-uniqueness of the states that achieve maximal departure
in the proposed scheme [Sec.~\ref{NoUnico}]. Explicitly, here the states can
be taken as $\rho _{A}=(1-2|z|)\varrho _{ab}+|z|(\mathrm{I}_{2}\otimes
|+\rangle \langle +|)$ and $\rho _{B}=(1-2|z|)\varrho _{ab}+|z|(\mathrm{I}%
_{2}\otimes |-\rangle \langle -|),$ where $\varrho _{ab}$ is an arbitrary
density matrix for two qubits. Hence, $\rho _{A}-\rho _{B}$ (jointly with $%
\rho _{B}-\rho _{A}$) recovers Eq.~(\ref{Delta}) with $x=-y=z.$

\section{Summary and Conclusions}

We have introduced an alternative distinguishability measure between quantum
states. In contrast to the standard definition based on maximization over
projectors, the proposed measure relies on a maximization over states [Eq.~(%
\ref{DrhoDefinition})]. This operation can be explicitly performed [Eq.~(\ref%
{DRhoEigenSol})], which allowed us to demonstrate that the proposed measure
is a metric in the space of density matrices based on an
operator-infinite-norm. In addition, it was shown that properties such as
convexity, monotonicity in bipartite Hilbert spaces, and invariance under
unitary transformations are also fulfilled.

Similarly to the usual projector-based definition, different operational
interpretations of the proposed distinguishability measure have been
established. It can be read as a maximization over states of a distance
between probabilities, each one being associated to each input state [Eq.~(%
\ref{MaxClassical})]. The distinguishability measure also defines the
probability of success in a hypothesis testing scenario [Eq.~(\ref{Testing}%
)] where a state is guessed in terms of a measurement process consisting of
a 1-rank projector and its complement [Eq.~(\ref{Unidimensional})].

The projector- and state-based definitions are equal when the Hilbert space
dimension is two or three [Eq.~(\ref{23})]. For higher dimensional spaces
[Eq.~(\ref{Constraints})] the relationship between both objects depends on
the eigenvalues of the difference of states. When the eigenvalue with
maximal absolute value is not degenerate, both measures coincide. When this
eigenvalue has maximal degeneracy, the state-based definition achieves its
minimal value with respect to the projector-based definition.

In contrast to other distances in Hilbert space, we demonstrated that the
proposed measure is able to quantify the quantum character of dissipative
open system dynamics. This result relies on the contractivity properties of
the proposed measure. For unital maps, contractivity is always satisfied
while, for non-unital maps, violation of contractivity is expected, meaning
that there could be states such that their distance increases after
application of the map. It was shown that maximal violation of contractivity
is always achieved when expanding the map to an extra ancillary Hilbert
space without dynamics [Eqs.~(\ref{ExtendedMap}) and~(\ref{EqualMax})]. The
quantumness measure for non-unital maps is defined by the constant
associated to this scheme which, in turn, can be written in terms of the
proposed distinguishability measure [Eqs.~(\ref{Mq}) and (\ref{Mq_Dinfinita}%
)].

We have studied some particular cases and examples that sustain the main
results and conclusions. The proposed measure may find applications in
quantum information tasks as well as in the characterization of open quantum
system dynamics. In particular, given that dissipative non-classical
(quantum) system-environment interactions lead to non-unital dynamics, the
present measure plays a central role when characterizing this
quantum-classical border.

\section*{Acknowledgments}

A.A.B. thanks fruitful discussions with Prof. Rolando Rebolledo as well as
financial support from Consejo Nacional de Investigaciones Cient\'{\i}ficas
y T\'{e}cnicas CONICET, Argentina. M.F.S thanks the financial support of
CNPq Project 302872/2019-1 and FAPERJ Project CNE - E-26/200.307/2023.

\appendix

\section{Maximization over states of an operator expectation value \label%
{MaxAverage}}

Let $A$ be an arbitrary Hermitian operator, $A=A^{\dagger }.$ Define its
maximized expectation value by%
\begin{equation}
\langle A\rangle _{\max }\equiv \max_{\{\rho \}}\left\vert \mathrm{Tr}[\rho
A]\right\vert ,  \label{Amax}
\end{equation}%
where the maximization is performed over positive definite normalized
density matrices, $\mathrm{Tr}[\rho ]=1.$ Introducing the eigenbasis $%
\{|i\rangle \}$ of the operator $A,$%
\begin{equation}
A|i\rangle =\lambda _{i}|i\rangle ,
\end{equation}%
where $\{\lambda _{i}\}$ are the corresponding eigenvalues, it follows that%
\begin{equation}
\langle A\rangle _{\max }=\max_{\{\rho \}}\Big{|}\sum_{i}\lambda _{i}\langle
i|\rho |i\rangle \Big{|}=\max_{\{P_{i}\}}\Big{|}\sum_{i}\lambda _{i}P_{i}%
\Big{|}.
\end{equation}%
Here, $P_{i}\equiv \langle i|\rho |i\rangle ,$ $0\leq P_{i}\leq 1.$ Using
the triangular inequality $(|a+b|\leq |a|+|b|),$ it follows that%
\begin{equation}
\Big{|}\sum_{i}\lambda _{i}P_{i}\Big{|}\leq \sum_{i}|\lambda _{i}P_{i}|\leq
\max_{\{i\}}(|\lambda _{i}|)\sum_{i}P_{i},
\end{equation}%
where $\max_{\{i\}}(|\lambda _{i}|)$ is the eigenvalue of $A$ with maximal
absolute value. Using that $\sum_{i}P_{i}=1,$ we obtain%
\begin{equation}
\langle A\rangle _{\max }=\max_{\{i\}}(|\lambda _{i}|).  \label{LamMax}
\end{equation}%
Hence, $\langle A\rangle _{\max }$ is the eigenvalue of the operator $A$
with maximal absolute value. On the other hand, we notice that $\langle
A\rangle _{\max }=0\Leftrightarrow A=0.$ Both implications follow
straightforwardly from Eqs.~(\ref{Amax}) and~(\ref{LamMax}) respectively.

The state $\rho $ that achieves the maximal value in the definition~(\ref%
{Amax}) can always be chosen as $\rho =|i_{\max }\rangle \langle i_{\max }|,$
where $|i_{\max }\rangle $ is the eigenstate associated to $%
\max_{\{i\}}(|\lambda _{i}|).$ If this eigenvalue (with a given sign) is
degenerate, $\rho $ can be taken as an arbitrary mixed state over the
corresponding subspace. On the other hand, if there exists a subspace with
null eigenvalues, $\{\lambda _{k}=0\},$ the demonstration remains the same
because $\sum_{i}P_{i}$ can always be normalized to one on the subspace with
non-null eigenvalues.

\section{General properties of $D_{\protect\rho }$ \label{PropertiesD}}

The distinguishability measure $D_{\rho }(\rho _{A},\rho _{B})$ fulfills
some general properties whose formulation and demonstration are provided
below.

$\blacksquare $ $D_{\rho }(\rho _{A},\rho _{B})$ is positive and bounded, 
\begin{equation}
0\leq D_{\rho }(\rho _{A},\rho _{B})\leq 1.
\end{equation}%
This results follows from the Eq.~(\ref{DRhoEigenSol}) after noticing that
Eq.~(\ref{Eigen}) implies that $\langle i|(\rho _{A}-\rho _{B})|i\rangle
=\langle i|\rho _{A}|i\rangle -\langle i|\rho _{B}|i\rangle =\zeta _{i},$
which is a difference between two populations leading to $-1\leq \zeta
_{i}\leq 1.$

$\blacksquare $ $D_{\rho }(\rho _{A},\rho _{B})$ is null if and only if $%
\rho _{A}=\rho _{B},$%
\begin{equation}
D_{\rho }(\rho _{A},\rho _{B})=0\ \ \ \ \Leftrightarrow \ \ \ \ \rho
_{A}=\rho _{B}.
\end{equation}%
Both implications follow from Eqs.~(\ref{DrhoDefinition}) and~(\ref%
{DRhoEigenSol}).

$\blacksquare $ $D_{\rho }(\rho _{A},\rho _{B})$ is a distance or \textit{%
metric} in the space of density operators, that is, in addition it satisfies,%
\begin{equation}
D_{\rho }(\rho _{A},\rho _{C})\leq D_{\rho }(\rho _{A},\rho _{B})+D_{\rho
}(\rho _{B},\rho _{C}),
\end{equation}%
the triangular inequality.

\textit{Demonstration}: By its definition [Eq.~(\ref{DrhoDefinition})] there
exists a state $\rho ^{\max }$ such that%
\begin{eqnarray*}
D_{\rho }(\rho _{A},\rho _{C}) &=&|\mathrm{Tr}[\rho ^{\max }(\rho _{A}-\rho
_{C})]| \\
\!\! &=&|\mathrm{Tr}[\rho ^{\max }(\rho _{A}-\rho _{B})]+\mathrm{Tr}[\rho
^{\max }(\rho _{B}-\rho _{C})]| \\
\!\! &\leq &\!\!\!|\mathrm{Tr}[\rho ^{\max }(\rho _{A}-\rho _{B})]|+|\mathrm{%
Tr}[\rho ^{\max }(\rho _{B}-\rho _{C})]| \\
\!\! &\leq &D_{\rho }(\rho _{A},\rho _{B})+D_{\rho }(\rho _{B},\rho _{C}),
\end{eqnarray*}%
establishing that $D_{\rho }(\rho _{A},\rho _{B})$ is a metric.\ The
inequality in the third line relies on the usual triangular inequality $%
(|a+b|\leq |a|+|b|).$

$\blacksquare $ Given a set of positive normalized weights, $%
\sum_{i}p_{i}=1, $ \textit{convexity} is%
\begin{equation}
D_{\rho }(\sum_{i}p_{i}\rho _{i},\sum_{i}p_{i}\sigma _{i})\leq
\sum_{i}p_{i}D_{\rho }(\rho _{i},\sigma _{i}),  \label{C1}
\end{equation}%
where the sets of states $\{\rho _{i}\}$ and $\{\sigma _{i}\}$ are arbitrary
ones. In the case in which $\sigma _{i}\rightarrow \sigma ,$ it follows%
\begin{equation}
D_{\rho }(\sum_{i}p_{i}\rho _{i},\sigma )\leq \sum_{i}p_{i}D_{\rho }(\rho
_{i},\sigma ).  \label{C2}
\end{equation}%
Thus, $D_{\rho }$ is convex in both entries.

\textit{Demonstration}: By its definition there exists a state $\rho ^{\max
} $ such that%
\begin{eqnarray*}
D_{\rho }(\sum_{i}p_{i}\rho _{i},\sum_{i}p_{i}\sigma _{i}) &=&|\mathrm{Tr}%
[\rho ^{\max }\sum_{i}p_{i}(\rho _{i}-\sigma _{i})]| \\
&=&|\sum_{i}p_{i}\mathrm{Tr}[\rho ^{\max }(\rho _{i}-\sigma _{i})]| \\
&\leq &\sum_{i}p_{i}|\mathrm{Tr}[\rho ^{\max }(\rho _{i}-\sigma _{i})]| \\
&\leq &\sum_{i}p_{i}\max_{\{\rho \}}|\mathrm{Tr}[\rho (\rho _{i}-\sigma
_{i})]| \\
&=&\sum_{i}p_{i}D_{\rho }(\rho _{i},\sigma _{i}),
\end{eqnarray*}%
where the triangular inequality was used in the third line. The
demonstration of Eq.~(\ref{C2}) is the same that before, replacing $\sigma
_{i}\rightarrow \sigma $ and using that $\sum_{i}p_{i}=1.$

$\blacksquare $ In a bipartite Hilbert space with subparts $a$ and $b,$ 
\textit{monotonicity }is%
\begin{equation}
D_{\rho }(\rho _{a},\sigma _{a})\leq D_{\rho }(\rho _{ab},\sigma _{ab}),
\end{equation}%
where $\rho _{a}=\mathrm{Tr}_{b}[\rho _{ab}]$ and $\sigma _{a}=\mathrm{Tr}%
_{b}[\sigma _{ab}].$

\textit{Demonstration}: there exists a state $\rho _{a}^{\max }$ that leads
to maximization,%
\begin{eqnarray*}
D_{\rho }(\rho _{a},\sigma _{a}) &=&|\mathrm{Tr}_{a}[\rho _{a}^{\max }(\rho
_{a}-\sigma _{a})]| \\
&=&|\mathrm{Tr}_{ab}[(\rho _{a}^{\max }\otimes \mathrm{I}_{b})(\rho
_{ab}-\sigma _{ab})]| \\
&\leq &\max_{\{\rho \}}|\mathrm{Tr}_{ab}[\rho (\rho _{ab}-\sigma _{ab})]| \\
&=&D_{\rho }(\rho _{ab},\sigma _{ab}),
\end{eqnarray*}%
where consistently $\rho $ (in the third line) is an arbitrary bipartite
state.

$\blacksquare $ \textit{Invariance under unitary rotations,}%
\begin{equation}
D_{\rho }(U\rho _{A}U^{\dag },U\rho _{B}U^{\dag })=D_{\rho }(\rho _{A},\rho
_{B}),
\end{equation}%
where $UU^{\dag }=\mathrm{I}.$ This result follows straightforwardly from
the definition~(\ref{DrhoDefinition}) after using the cyclic property of the
trace and noting that $U^{\dag }\rho U$ is also an arbitrary state.

\section{Contractivity of the zero temperature qubit map\label{ThermalD2}}

Given two arbitrary states $\rho _{A}$ and $\rho _{B}$ their difference is
denoted as (see also Appendix~\ref{Space})%
\begin{equation}
\Delta \equiv \rho _{A}-\rho _{B}=\left( 
\begin{array}{cc}
\delta p & \delta c \\ 
\delta c^{\ast } & -\delta p%
\end{array}%
\right) .
\end{equation}%
The eigenvalues of $\Delta $ are $\pm \sqrt{\delta p^{2}+|\delta c|^{2}}.$
Consequently,%
\begin{equation}
D_{\rho }[\Delta ]=\sqrt{\delta p^{2}+|\delta c|^{2}}.  \label{DInput2}
\end{equation}%
The action of the map on the difference of states $\Delta ,$ from Eq.~(\ref%
{ThermalMap}), is%
\begin{equation}
\mathcal{E}[\Delta ]=\left( 
\begin{array}{cc}
(1-\gamma )\delta p & \sqrt{1-\gamma }\delta c \\ 
\sqrt{1-\gamma }\delta c^{\ast } & -(1-\gamma )\delta p%
\end{array}%
\right) .
\end{equation}%
The eigenvalues of $\mathcal{E}[\Delta ]$ are $\pm \sqrt{(1-\gamma
)^{2}\delta p^{2}+(1-\gamma )|\delta c|^{2}}.$ Consequently,%
\begin{equation}
D_{\rho }[\mathcal{E}[\Delta ]]=\sqrt{(1-\gamma )}\ \sqrt{(1-\gamma )\delta
p^{2}+|\delta c|^{2}}.  \label{DOutput2}
\end{equation}%
From Eqs.~(\ref{DInput2}) and (\ref{DOutput2}) it follows that%
\begin{equation}
D_{\rho }[\mathcal{E}[\Delta ]]\leq D_{\rho }[\Delta ].
\end{equation}%
As expected, usual contractivity [Eq.~(\ref{ContractorRho})] is fulfilled
for any input state.

\section{Space associated to difference of quantum states\label{Space}}

Here we establish how to parametrize in a general way the difference between
two density matrices. Given two states $\rho _{A}$ and $\rho _{B}$ define%
\begin{equation}
\Delta \equiv \rho _{A}-\rho _{B}.
\end{equation}%
Hence, instead of $\rho _{A}$ and $\rho _{B}$, the goal is to parametrize $%
\Delta $ in an independent way. Written in terms of the eigensystem $(\rho
_{A}-\rho _{B})|i\rangle =\zeta _{i}|i\rangle ,$ it follows%
\begin{equation}
\Delta =\sum_{i}\zeta _{i}|i\rangle \langle i|.
\end{equation}%
Thus, $\Delta $ can be characterized in terms of an arbitrary basis $%
\{|i\rangle \}$\ and the eigenvalues $\{\zeta _{i}\}.$ Given that $\mathrm{Tr%
}[\Delta ]=0,$ the addition of the eigenvalues must vanish. Furthermore,
each eigenvalue must be in the interval $[-1,1],$ that is,%
\begin{equation}
|\zeta _{i}|\leq 1,\ \ \ \ \ \ \ \ \sum_{i=1}^{\dim (\emph{H})}\zeta _{i}=0,
\label{ModuloEigen}
\end{equation}
where $\dim (\emph{H})$\ is the dimension of the Hilbert space. Added to
these conditions, the sum of an arbitrary number of eigenvalues also must be
in the interval $[-1,1].$ This condition can be explicitly written by
introducing the vector of eigenvalues $\mathbf{\zeta }=(\zeta _{1},\zeta
_{1},\cdots ,\zeta _{n}),$ and the vector $\mathbf{b}=(b_{1},b_{2},\cdots
,b_{n}),$ whose components are $b_{i}=0$ or $b_{i}=1.$ Thus, it must be
satisfied that for all vectors $\mathbf{b}$ $[\mathbf{b}\neq (1,1,\cdots
,1)] $ that 
\begin{figure}[t]
\includegraphics[bb=0 0 380 380,angle=0,width=6cm]{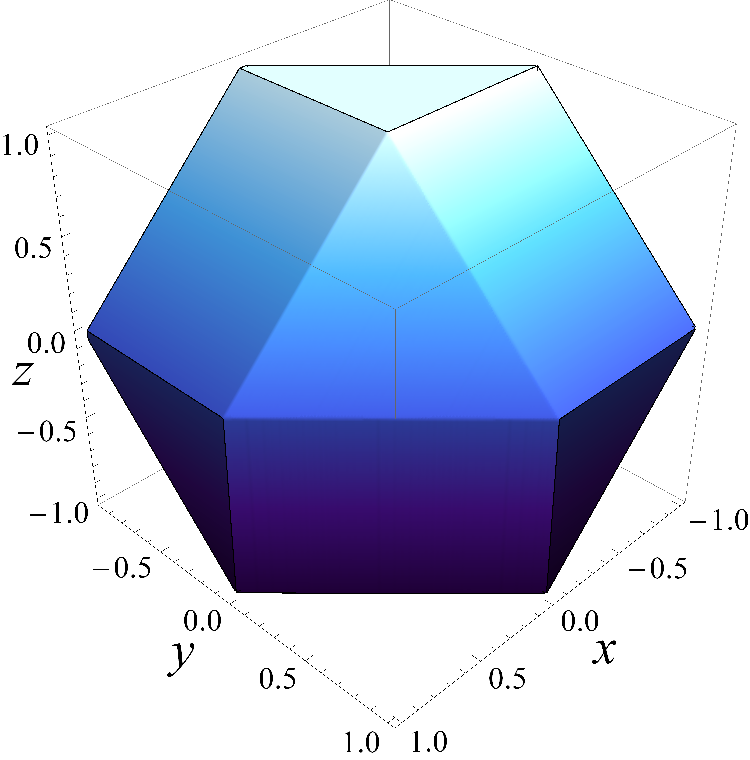}
\caption{Domain of the parameters $(x,y,z)$ that set the eigenvalues of the
difference of states $\Delta =\protect\rho _{A}-\protect\rho _{B}$ defined
by Eq.~(\protect\ref{DeltaXYZ}) (four dimensional Hilbert space). The
constraints on $(x,y,z)$ are defined by Eqs.~(\protect\ref{Volume1}) and~(%
\protect\ref{Volume2}).}
\label{Solid3D}
\end{figure}
\begin{equation}
|\mathbf{\zeta .b}|=\Big{|}\sum_{k=1}^{\dim (\emph{H})}\zeta _{k}b_{k}\Big{|}%
\leq 1.  \label{SumaEigen}
\end{equation}%
We notice that the condition $|\zeta _{i}|\leq 1$ is recovered when $\mathbf{%
b}$ is the canonical basis, $b_{k}=\delta _{ki}.$ On the other hand, the
condition $\sum_{i=1}^{\dim (\emph{H})}\zeta _{i}=0$ can be written as $|%
\mathbf{\zeta .b}|=0$ where $\mathbf{b}=(1,1,\cdots ,1).$

\textit{Demonstration}: The previous conditions [Eq.~(\ref{ModuloEigen})
and~(\ref{SumaEigen})] can be derived as follows. Straightforwardly, the
condition $\mathrm{Tr}[\Delta ]=0$ implies that the eigenvalues fulfill $%
\sum_{i}\zeta _{i}=0.$ On the other hand, given $\Delta ,$ there must exist
states $\rho $ and $\sigma $ such that $\Delta +\sigma =\rho .$ Taking
matrix elements in the basis $\{|i\rangle \}$ associated to $\Delta $ it
follows that $\langle i|\Delta |i\rangle +\langle i|\sigma |i\rangle
=\langle i|\rho |i\rangle .$ Given that $0\leq \langle i|\rho |i\rangle \leq
1$ and $0\leq \langle i|\sigma |i\rangle \leq 1,$ using that $\langle
i|\Delta |i\rangle =\zeta _{i},$ it follows that $|\zeta _{i}|\leq 1$ [Eq.~(%
\ref{ModuloEigen})]. Furthermore, the addition of an arbitrary number of
diagonal components must be less than one. For example, $\langle i|\Delta
|i\rangle +\langle k|\Delta |k\rangle +\langle i|\sigma |i\rangle +\langle
k|\sigma |k\rangle =\langle i|\rho |i\rangle +\langle k|\rho |k\rangle \leq
1.$ In general, $\sum_{k=1}^{\dim (\emph{H})}b_{k}\langle k|\Delta |k\rangle
+\sum_{k=1}^{\dim (\emph{H})}b_{k}\langle k|\sigma |k\rangle
=\sum_{k=1}^{\dim (\emph{H})}b_{k}\langle k|\rho |k\rangle \leq 1,$ which
leads to Eq.~(\ref{SumaEigen}).

\textit{Four dimensional case}: Below, we characterize the difference of
states $\Delta $ in a four dimensional space. The basis where it is diagonal
remains unspecified. Thus, we write%
\begin{equation}
\Delta =\rho _{A}-\rho _{B}=\mathrm{diag}\{x,y,z,-(x+y+z)\}.
\label{DeltaXYZ}
\end{equation}%
The condition $\mathrm{Tr}[\Delta ]=0$ is automatically fulfilled.
Furthermore, the condition Eq.~(\ref{ModuloEigen}) on the eigenvalues, here
denoted as $x,$ $y,$ $z,$ and $-(x+y+z),$ is satisfied under the conditions%
\begin{equation}
|x|\leq 1,\ \ \ \ |y|\leq 1,\ \ \ \ |z|\leq 1,\ \ \ \ |x+y+z|\leq 1.
\label{Volume1}
\end{equation}%
In addition, Eq.~(\ref{SumaEigen}) leads to the extra constraints%
\begin{equation}
|x+y|\leq 1,\ \ \ \ |x+z|\leq 1,\ \ \ \ |y+z|\leq 1.  \label{Volume2}
\end{equation}%
The inequalities Eqs.~(\ref{Volume1}) and (\ref{Volume2}), in the space
defined by $(x,y,z),$ define a 3-dimensional body with fourteen faces. It is
plotted in Fig.~\ref{Solid3D}.


\begin{thebibliography}{99}
\bibitem{POVM} Let $E=\{E_{m}\}$ be a set of positive operator value
measure, $\sum_{m}E_{m}=\mathrm{I}.$ Define $p_{A}^{(m)}\equiv \mathrm{Tr}%
[E_{m}\rho _{A}]$ and $p_{B}^{(m)}\equiv \mathrm{Tr}[E_{m}\rho _{B}].$ Then, 
$D_{\Pi }(\rho _{A},\rho _{B})=\max_{\{E\}}D_{c}(p_{A},p_{B}),$ where the
(classical-incoherent) probability distance is $D_{c}(p_{A},p_{B})\equiv
(1/2)\sum_{m}|p_{A}^{(m)}-p_{B}^{(m)}|.$
\bibitem{ceros} If there are $n_{0}$ null eigenvalues the argument remains
the same under the replacement $\dim (\mathcal{H})\rightarrow \dim (\mathcal{%
H})-n_{0}.$
\bibitem{Entero} Assume that $\max_{\{i\}}\{|\zeta _{i}|\}=\zeta _{\max
}^{(+)}$ is positive. It maximal degeneracy occurs when\ $\zeta
_{i}^{(+)}=\zeta _{\max }^{(+)}$ $\forall i.$ Using that $%
\sum_{i=1}^{n_{+}}\zeta _{i}^{(+)}=\sum_{j=1}^{n_{-}}|\zeta _{i}^{(-)}|,$ it
follows that $\zeta _{\max }^{(+)}n^{+}=\sum_{j=1}^{n_{-}}|\zeta
_{i}^{(-)}|\leq \zeta _{\max }^{(+)}n^{-},$ implying $n^{+}\leq n^{-}.$
Given that $n^{+}+n^{-}=\dim (\mathcal{H}),$ it is fulfilled that $%
2n^{+}\leq \dim (\mathcal{H}).$ Hence, the maximal value of $n^{+}$ is $%
\mathcal{N}=\mathrm{Int}[\dim (\mathcal{H})/2].$ The same reasoning applies
when $\max_{\{i\}}\{|\zeta _{i}|\}=|\zeta _{\max }^{(-)}|$ is negative.
\end{thebibliography}
\end{document}